\documentclass[prb,twocolumn,superscriptaddress,floatfix]{revtex4}
\usepackage{epsfig}
\usepackage{epstopdf}

\begin{document}

\title{The role of FCC tetrahedral subunits in the phase behavior of medium sized Lennard-Jones clusters.}

\author{Ivan Saika-Voivod}
\email{saika@mun.ca}
\affiliation{ 
Department of Physics and Physical Oceanography, Memorial University of Newfoundland, St. JohnÕs, NL, A1B 3X7, Canada}

\author{Louis Poon}
\affiliation{Department of Chemistry, University of Saskatchewan,
Saskatoon, Saskatchewan, S7N 5C9, Canada}

\author{Richard K. Bowles}
\email{richard.bowles@usask.ca}
\affiliation{Department of Chemistry, University of Saskatchewan, Saskatoon, Saskatchewan, S7N 5C9, Canada}

\date{\today}

\begin{abstract}
The free energy of a 600-atom Lennard-Jones cluster is calculated as a function of surface and bulk crystallinity in order to study the structural transformations that occur in the core of medium sized clusters.  Within the order parameter range studied, we find the existence of two free energy minima at temperatures near freezing. One minimum, at low values of both bulk and surface order, belongs to the liquid phase. The second minimum exhibits a highly ordered core with a disordered surface and is related to structures containing a single FCC-tetrahedral subunit, with an edge length of seven atoms ($l=7$), located in the particle core. At lower temperatures, a third minimum appears at intermediate values of the bulk order parameter which is shown to be related to the formation of multiple $l=6$ tetrahedra in the core of the cluster. We also use molecular dynamics simulations to follow a series of nucleation events and find that the clusters freeze to structures containing $l=5,6,7$ and $8$ sized tetrahedra as well as those containing no tetrahedral units. The structural correlations between bulk and surface order with the size of the tetrahedral units in the cluster core are examined. Finally, the relationships between the formation of FCC tetrahedral subunits in the core, the phase behavior of medium sized clusters and the nucleation of noncrystalline global structures such as icosahedra and decahedra are discussed.
\end{abstract}


\maketitle

\section{Introduction}

The interplay between surface and volume effects in nanoscale atomic clusters result in a wide range of unique structural and thermodynamic properties that are not observed in bulk systems.~\cite{bal05} The lowest energy configurations for small clusters tend to be the noncrystalline structures such as the MacKay iscosahedron ({\it Ih}) and Marks decahedron ({\it Dh}), with magic number clusters consisting of complete closed shell structures being very stable.~\cite{walesbook} As the cluster size increases, there must eventually be a crossover to the bulk face-centered-cubic ({\it FCC}) and hexagonal close packed ({\it HCP}) based crystals. However, extensive computer simulation based searches~\cite{walesbook, xia104} of the potential energy surfaces for Lennard-Jones (LJ) clusters suggest that the {\it Ih} and {\it Dh} structures remain the dominant lowest energy states for clusters containing up to 1000 atoms.~\cite{xia204}

Advanced simulation techniques, such as parallel tempering,~\cite{temp} have allowed accurate calculations of the free energy landscape for systems in the small cluster regime~\cite{lab90,che92,doy99,nei00,cal00,fra01,mand06}. For example, the 38-atom LJ cluster shows a phase transition between the MacKay icosahedral phase and the energetically more favorable, truncated octahedron structure~\cite{doy99,nei00,cal00}  while Mandelshtam et al~\cite{mand06} have recently reported simulation studies of the phase behavior of LJ clusters with up to 147 atoms. They found the caloric curves of these  cluster sizes exhibited two characteristic peaks. The low temperature peak corresponded to the melting of the MacKay surface layer, leading to an icosahedral core-ordered phase, and the high temperature peak was associated with the core melting to the liquid. Recently, Zhan et al~\cite{zhan07} showed that a combination of local and global structural order parameters, based on the Steinhardt measures,~\cite{steinhardt} could be used to distinguish between all the different structural transitions observed in small LJ clusters.

Very little is known about the phase behavior of medium sized LJ clusters. Doye and Calvo~\cite{doy02} developed a coarse grained approach to study the effects of temperature on the relative stability of the {\it Ih}, {\it Dh} and {\it FCC} structures of larger clusters and found that the noncrystalline states remain the most stable up to cluster sizes in the order of many hundreds of thousands of LJ atoms. The 
 309-atom LJ cluster has been shown to exhibit a variety of structural transformations involving major rearrangements of the core atoms in addition to the surface reconstruction type transitions observed in smaller clusters.~\cite{noya06}  These include core structures based on the formation of twinned {\it FCC} tetrahedra with five-atom edges and isolated {\it FCC} tetrahedra with six-atom edges.
Polak~\cite{pol08} recently studied the internal structure of LJ clusters by cooling clusters in the size range $N=55-923$ to $T=0.05$ from the liquid state and using the coordination polyhedron method~\cite{pol03} which identifies an individual atom as having the local environment consistent with an FCC, HCP or Ih symmetry etc. He classified the resulting clusters as belonging to a variety of structural families depending on the presence and arrangement of atoms with five fold symmetry (Ih) or the appearance of defective crystal states. However, it remains a significant challenge to understand how these structures and structure families are related to the phase behavior or nucleation pathways of the clusters.

The goal of the present paper is to study the structural transformations occuring in the core of intermediate sized Lennard-Jones clusters. To this end, we calculate the free energy surface of a 600-atom cluster over a range of temperatures near the freezing transition as a function of order parameters that measure the core and surface order of the cluster. The resulting free energy surface exhibits minima associated with the liquid, a core ordered structure consisting of a seven-atom {\it FCC} tetrahedron surrounded by a disordered surface and  structures with multiple twinned six-atom edge tetrahedra.  We also employ molecular dynamics (MD) simulations to understand how the phase behavior of the system influences the type of kinetic structures observed following a nucleation event. The remainder of the paper is organized as follows: Section II provides the details of our free energy calculations, MD simulations and the use of inherent structure quenches and common neighbor analysis to determine the particle structures. Section III contains our results and discussion while the conclusions are contained in Section IV.

\section{Simulation Methods}

Our system consists of $600$ particles of mass $m$ interacting with the Lennard-Jones 12-6 pair potential,
\begin{equation}
U(r)=4\epsilon\left[\left(\sigma/r\right)^{12}-\left(\sigma/r\right)^6\right]\mbox{ ,}\\
\label{pot}
\end{equation}
where $r$ is the distance between two particles, $\epsilon$ sets the energy scale and $\sigma$ the length scale. 
All quantities are reported here in reduced units, with unit time given by $\sqrt{m \sigma^2/ \epsilon}$.
We employ a cubic simulation cell with periodic boundary conditions and volume  $V=30^3$, which is large enough to ensure the clusters do not interact with their periodic images, but small enough to stabilize the liquid drop with respect to evaporation. 
One or two particles can be found detached from the cluster at times for the range of $T$ we study.

The free energy is calculated as a function of both bulk and surface crystallinity order parameters,
$Q_{\rm b}$ and $Q_{\rm s}$ respectively,  
both of which are based on the Steinhardt~\cite{steinhardt} bond order parameter, $Q_6$, made popular for
simulation studies of crystallization by Frenkel and coworkers~\cite{frenkel1996}.
The difference is that $Q_{\rm s}$ is determined by considering only surface atoms, and $Q_{\rm b}$ by using only bulk atoms. They are defined as follows:
\begin{equation}
Q_{\rm b,s}=\left[ \frac{4\pi}{13} \sum_{m=-6}^6 \left|  
\frac{1}{N_{\rm b,s}} \sum_{i=1}^{N_{\rm b,s}} q_{6m}(i)
\right|^2 \right]^{1/2}\mbox{,}
\label{def_q}
\end{equation}
where
\begin{equation}
q_{6m}(i)=\frac{1}{n(i)}\sum_{j=1}^{n(i)} Y_{6m}(\hat r_{ij}),\\
\end{equation}
the second sum in Eq.~\ref{def_q} over $i$ is over all $N_{\rm b}$ bulk or $N_{\rm s}$ surface particles,
$n(i)$ is the number of neighbors for particle $i$,
$Y_{6m}(\theta,\phi)$ are spherical harmonic functions, 
and $\hat r_{ij}$ is the unit vector pointing from particle $i$ to a neighbor and specifying
the elevation and azimuth angles that their bond makes with respect to the coordinate system of the simulation cell.
Two particles are considered to be neighbors if the distance between them is less than or equal
to $1.363$ and this distance is chosen to minimize the sensitivity of $Q_{\rm b,s}$ to 
this parameter. Surface particles are distinguished from bulk particles by means of a slightly modified version of the ``cone'' algorithm.~\cite{cone} Our cone is parameterized by an apex angle of $120^\circ$, a slant length of $1.5$ and it has a spherically rounded bottom. For a $600$ particle cluster, about $40\%$ of the particles belong to the surface. The value of $Q_{\rm s}=0.20729$ for a perfect MacKay icosahedron.~\cite{cone}

We then obtain the free energy,
\begin{equation}
F(Q_{\rm b},Q_{\rm s})=-k_{\rm B}T\ln{P(Q_{\rm b},Q_{\rm s})}+{\rm const},\\
\label{eq:fe}
\end{equation}
over a range of $T$, with $k_{\rm B}$ being Boltzmann's constant.  $P(Q_{\rm b},Q_{\rm s})$ is the probability of observing the equilibrium system with given values of $Q_{\rm b}$ and $Q_{\rm s}$.
To determine $P$, we cover a domain ranging from $0.045$ to $0.17$ in $Q_{\rm b}$ 
and from $0.04$ to $0.135$ in $Q_{\rm s}$
with  48 overlapping
$4\Delta q \times 4\Delta q$ square subdomains, or ``patches'', where $\Delta q = 0.005$.
For each patch, we build up a $4 \times 4$ histogram that yields $P$
up to a multiplicative constant on the patch domain. 

To simulate the system on a patch, we employ constrained Monte Carlo (MC), biasing both $Q_{\rm b}$ and $Q_{\rm s}$. In our scheme, a MC move is rejected if it violates either
$| Q_{\rm b} - Q_{\rm b}^k | \le 2 \Delta q$ or
$| Q_{\rm s} - Q_{\rm s}^k | \le 2 \Delta q$,
where 
$Q_{\rm b}^k$ and $Q_{\rm s}^k$ mark the center of the patch.
One MC step consists of 600 single particle MC moves.
Using the self-consistent histogram
method,~\cite{frenkelbook} we then match the histograms to obtain
$P$, up to an overall normalization constant, for all $Q_{\rm b}$ and $Q_{\rm s}$.

To facilitate equilibration, we run all the patches in parallel, allowing configurations to switch between neighboring patches.
Two neighboring patches overlap along one entire edge by $\Delta q$, with up to four neighbors per patch.
Random attempts to switch configurations between two neighboring patches
occur every 236 MC steps on average.
Switches are accepted if configurations from both patches lie in the overlap region.
Additionally, we simultaneously simulate four grids at different $T$.
Each patch, therefore, has an additional one or two neighbors, in $T$, with
which it can attempt configuration switches.  Random $T$-switches are attempted between patches
in neighboring grids on average every 70 MC steps, and are accepted according to the standard
parallel tempering acceptance criteria.~\cite{frenkelbook}  With the four $T$ grids, there are 192
patches running in parallel.

The simulation grids are seeded with configurations from unconstrained MC runs.
Several sets of $T$ are used to equilibrate the system and to find a suitable $T$ range over which 
the liquid freezes to structures within our range of $Q_{\rm b}$ and $Q_{\rm s}$.  
The four $T$ reported in this work are
$0.475, 0.480, 0.485$ and $0.490$. To help quantify how the system equilibrates, we measure the decorrelation of bonds between neighbors. At the lowest temperature, $T=0.475$, liquid-like configurations show large
decorrelation, with only $7\%$ of initial bonds remaining after 60,000 MC steps.  Crystal-like configurations
show less decorrelation, with about $50\%$ of bonds remaining.  For $T=0.490$, $7\%$ of initial bonds
remain on average for all configurations.  After initial equilibration, the system is run for 6,760,000
MC steps in each patch. That is, our free energy surfaces are determined over approximately $2\times 10^{11}$ single particle MC moves at each temperature.

Despite the rather lengthy parallel runs (tempered in three dimensions), the free energy surfaces do continue to evolve slowly over time.
While the minima in our free energy surfaces appear to be stable, the potential energy plotted
as a function of MC steps for patches near saddle points shows some drift.
The slow nature of the drift, as well as the correspondence between our MC results and 
the molecular dynamics (MD) freezing runs, as discussed below, give us some confidence as to the qualitative features of the 
free energy surface, if not the precise location and heights of saddle points.  
It is acknowledged that it is a formidable computational task to fully equilibrate clusters 
of hundreds of particles~\cite{noya06,mand06}.



\begin{figure}
\hbox to \hsize{\epsfxsize=1\hsize\hfil\epsfbox{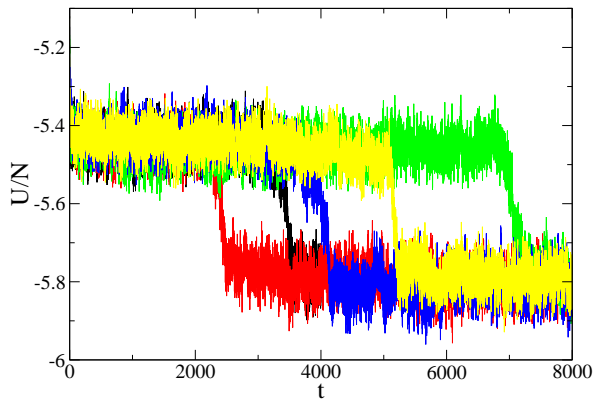}}
\caption{The potential energy of the cluster per particle as a function of time for five individual MD runs. (In color online.) }
\label{fig:mdet}
\end{figure}

We also perform a 
total of $100$ MD simulations in the $N,V,T$ ensemble using the Verlet algorithm, with a time step in reduced units of $\Delta t=0.001$, to integrate the equations of motion. The Anderson thermostat is used to control the temperature. The starting configurations for these runs are obtained from a single equilibrium simulation of a cluster performed at $T=0.53$ by saving configurations every $20000$ time steps after an initial equilibrium period of $2\times 10^6$ time steps. At the start of each run, the temperature is instantaneously decreased to $T=0.44$, which is well below the approximate freezing temperature for a cluster of this size~\cite{pol08}, and the trajectory followed for a time of $t=8000$. Fig.~\ref{fig:mdet} shows the potential energy of the cluster per particle, $U/N$, as a function of time for several MD trajectories. The energy of the starting configuration, obtained at $T=0.53$, is above $U/N\approx-5.2$ but this rapidly decays to approximately $-5.42$ as the liquid cluster establishes its metastable equilibrium before nucleating to a lower energy state with $U/N\approx -5.8$. $92\%$ of the runs nucleate while eight runs remain liquid on the time scale of the simulation. Molecular dynamics simulations of the clusters at higher temperatures do not show a significant number of nucleation events. 

The structural properties of a cluster are studied by performing a conjugate gradient quench of a configuration to its local potential energy minimum or inherent structure.~\cite{ih} This effectively removes thermal noise and enhances the signal obtained from our structural analysis. A common neighbor analysis~\cite{cla93,hen01} (CNA) is used to provide us with information concerning the structure of each individual particle based on its local environment and shared neighbors. Bulk particles can be identified as FCC,  HCP, icosahedral or amorphous. Surface particles are classified as belonging to the 111 or 100 surface,  an FCC edge, or an iscosahedral  edge, join or vertex, or finally as an amorphous surface atom. This approach has been used previously in the study of melting and freezing in small clusters.~\cite{hen01,noya06} It is worth noting that the criteria for identifying surface atoms in CNA is based solely on the number of neighbors, which differs from the cone analysis used in our $Q_{\rm b,s}$ work and results in  $37\%$ of the atoms in the cluster appearing on the surface. This analysis is applied to configurations obtained from the free energy simulations and to the final configurations of the MD simulations.

\section{Results and Discussion}

\begin{figure}
\hbox to \hsize{\epsfxsize=1\hsize\hfil\epsfbox{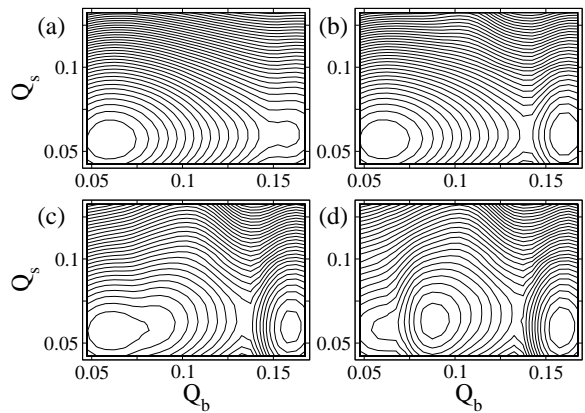}}
\caption{Contour plots of the free energy, $F(Q_{\rm b},Q_{\rm s})/kT$ as a function of $Q_{\rm b}$ and $Q_{\rm s}$ at (a) $T=0.490$, (b) $T=0.485$, (c) $T=0.480$ and (d) $T=0.475$. The contour lines are in increments  of $0.5/kT$.}
\label{fig:fe}
\end{figure}

\begin{figure}
\hbox to \hsize{\epsfxsize=1\hsize\hfil\epsfbox{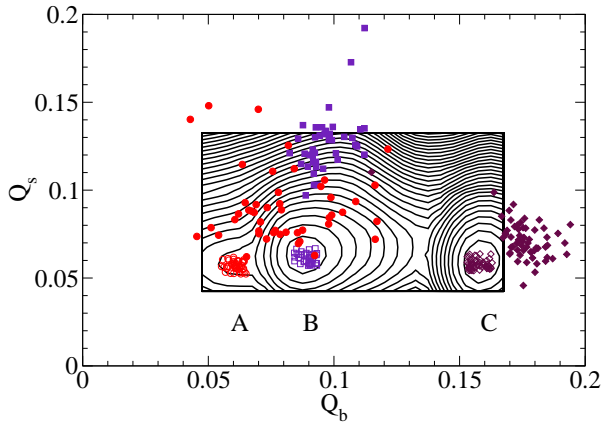}}
\caption{$Q_s$ as a function of $Q_b$ for the quenched inherent structures (filled symbols) obtained from starting configurations (open symbols) taken from regions A (circles), B (squares) and C (diamonds) on the free energy surface at $T=0.475$. The free energy contour plot from Fig.~\ref{fig:fe}d is included for reference. (In color online.) }
\label{fig:qmap}
\end{figure}

The free energy surfaces at the highest temperatures studied (Figs.~\ref{fig:fe}a and~\ref{fig:fe}b) have a free energy minimum centered at low values of $Q_b$ and $Q_s$, corresponding to the liquid, but we also see the appearance of a metastable state around $Q_b\approx0.16, Q_s\approx0.06$, which becomes deeper with decreasing temperature and eventually becomes more stable than the liquid. At $T=0.480$ (Fig.~\ref{fig:fe}c), we see the liquid minimum become distorted by the growth of a new minimum that is fully developed by $T=0.475$, at $Q_b\approx0.087, Q_s\approx0.07$ (Fig.~\ref{fig:fe}d). To understand the structure of the clusters associated with these locally stable free energy minima, we select a subset of the configurations from the $T=0.475$ free energy calculations with values of $Q_b$ and $Q_s$ that are consistent with the three minima, labeled A, B and C in Fig.~\ref{fig:qmap}, and quench them to their inherent structures using a conjugate gradient minimization of their potential energy. Configurations in group A have initial values of the order parameters consistent with the liquid minimum and the structures become broadly dispersed over a wide range of  $Q_b$ and $Q_s$ when quenched. The CNA ``signatures" of these inherent structures, listed in Table~\ref{tab:cna}, show that a majority of the particles in both the bulk and surface are amorphous, i.e. cannot be identified with any of the regular solid-like environments, and that the liquid clusters can be clearly distinguished from the solid clusters on this basis. Quenching the configurations from region B on the free energy surface leads to a significant increase in the surface order without much change in $Q_b$, while the configurations from region C order more in the bulk and remain relatively well grouped.

\begin{table}
  \begin{tabular}{@{} lccccccc @{}}
    \toprule
    & A & B & C& $l=5$& $l=6$& $l=7$ & $l=8$\\
    \hline
    N$_{\rm configs}$ & 47 & 40 & 58 & 6 & 6 & 24 & 5 \\
    Bulk FCC  & 18$\pm$9     & 70$\pm$9      & 100$\pm$11    & 36$\pm$4    &72$\pm$5         &101$\pm$4    &126$\pm$3\\ 
    Bulk HCP  & 53$\pm$36   & 180$\pm$5   & 166$\pm$22    & 191$\pm$4    & 185 $\pm$3    &175$\pm$5    &166$\pm$5\\ 
    Bulk ICO   & 99$\pm$16   & 77$\pm$10    & 52$\pm$5        & 111$\pm$9     &75$\pm$5  	 &53$\pm$4       &39$\pm$2\\ 
    Bulk Non   & 208$\pm$39 & 61$\pm$6    & 60$\pm$14      & 57$\pm$13     & 54 $\pm$6      &53$\pm$5       &50$\pm$6\\ 
    Surf 111    & 17$\pm$5     & 26$\pm$5    & 30$\pm$6         & 24 $\pm$4    & 33 $\pm$7        &39$\pm$8      &41$\pm$9\\ 
    Surf 100    & 12$\pm$4    & 28$\pm$4     & 34$\pm$ 6        & 20$\pm$ 3    & 24 $\pm$5       &29$\pm$3       &33$\pm$5\\ 
    FCC edge & 13$\pm$5     & 32$\pm$6     & 37$\pm$ 6        & 20$\pm$5    & 32 $\pm$8        &35$\pm$6       &40$\pm$3\\ 
    ICO vert    & 37$\pm$8     & 16$\pm$4     & 12$\pm$ 4        & 18$\pm$5    &  18 $\pm$4        &12$\pm$6      &11$\pm$4\\ 
    ICO edge  & 4$\pm$3       & 4$\pm$2       & 2$\pm$2           & 10$\pm$7      & 6 $\pm$5         &5$\pm$4        &2$\pm$3\\ 
    ICO join    & 8$\pm$5       & 20$\pm$7      & 20$\pm$5       & 22$\pm$10     & 22 $\pm$4        &23$\pm$5      &26$\pm$3\\ 
    Surf Non   &129$\pm$13  & 82$\pm$6     & 82$\pm$27      & 88$\pm$6     & 76 $\pm$5          &72$\pm$7      &64$\pm$8\\ 
    \hline  
  \end{tabular}
  \caption{The average number of atoms in different local environments obtained from the CNA of quenched configurations from regions A,  B and C of the free energy surface at $T=0.475$, and the $l=5,6,7,8$ clusters obtained in the MD freezing trajectories. The errors correspond to the standard deviation obtained from sample sizes of N$_{\rm configs}$.}
  \label{tab:cna}
  \end{table}

\begin{figure}
\hbox to \hsize{\epsfxsize=1\hsize\hfil\epsfbox{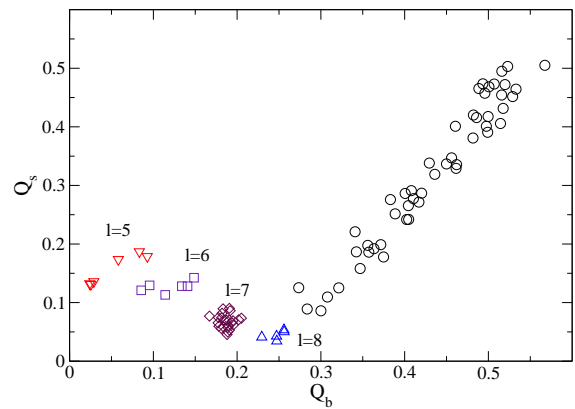}}
\caption{$Q_s$ as a function of $Q_b$ for the quenched inherent structures obtained from the MD runs. The clusters containing FCC tetrahedral subunits with edge lengths $l=5$ (down triangles), $l=6$ (squares), $l=7$ (diamonds) and $l=8$ (up triangles) are labeled respectively. The remaining structures (circles) contain no complete tetrahedral subunits.}
\label{fig:q6md}
\end{figure}

Fig.~\ref{fig:q6md}, which plots $Q_b$ vs  $Q_s$ for the quenched inherent structures from the MD simulations, shows that there are clear correlations between the two structural order parameters. Surface and bulk order appear to be negatively correlated at low values of $Q_b$, i.e. surface ordering improves at the expense of bulk order, but the two order parameters become positively correlated for $Q_b>0.26$.  We also note that there is a degree of clustering of the structures in parameter space with the largest grouping of $25$ structures being centered around $Q_b=0.19, Q_s=0.07$.

\begin{figure*}
\includegraphics[width=7in]{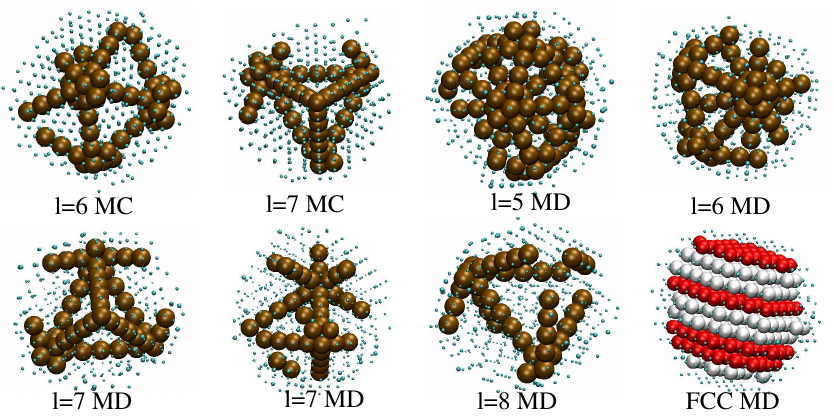}
\caption{ Representative inherent structures of the clusters from region B (l=6, MC) and C (l=7, MC) in the free energy plot in Fig.~\ref{fig:qmap} and regions $l=$5, 6, 7 and 8 from the MD simulations in Fig.~\ref{fig:q6md}. The bulk icosahedral atoms appear brown and the remaining atoms have been reduced in size and colored blue for clarity. The FCC, MD structure is representative of the structures formed in the MD runs that had no bulk icosahedral atoms. The bulk FCC and HCP atoms appear grey and red respectively and the remaining atoms have been reduced in size and colored blue for clarity. (In color online.) }
\label{fig:vmd}
\end{figure*}

\begin{figure}
\hbox to \hsize{\epsfxsize=1\hsize\hfil\epsfbox{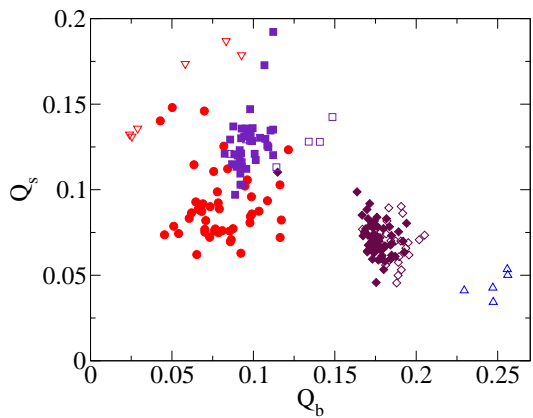}}
\caption{A comparison of $Q_s$ as a function of $Q_b$ for the quenched inherent structures from the free energy surface (region A (filled circles), B (filled squares) and C (filled diamonds)) with the MD simulations ($l=5$ (open down triangles), $l=6$ (open squares), $l=7$ (open diamonds) and $l=8$ (open up triangles)). 
}
\label{fig:qcomp}
\end{figure}

The direct visualization of the clusters from both the MC free energy calculations and the MD simulations reveal the appearance  of a series of structures based on the construction of FCC - tetrahedral subunits of different sizes. The edges of the tetrahedra are formed by the five-fold symmetric particles, i.e. the bulk icosahedral atoms, and we denote the size of a tetrahedron by the number of atoms along a single edge, $l$, including the vertex atoms at each end. The faces are formed from a plane of HCP atoms while the core of the tetrahedra are filled with FCC atoms. Fig.~\ref{fig:vmd} shows representative inherent structures that contain tetrahedra ranging from size $l=5$ through to $l=8$, although, in the last case, none of the clusters contains a complete tetrahedra. The structures containing different sized tetrahedral groups have distinct CNA signatures in the number of bulk FCC and bulk icosahedral atoms and have been labelled in Fig.~\ref{fig:q6md}.  It is clear that the clustering of structures with respect to the order parameters is due to the presence of the different subunits in the core of the particle. Also, direct visualization, along with a comparison of the CNA analysis and $Q_b,Q_s$ (see Fig.~\ref{fig:qcomp}) for the structures obtained from the free energy surfaces and MD runs, suggest that free energy minimum C, with the greatest amount of core order, is associated with formation of $l=7$ sized tetrahedra while the minimum B is related to structures containing $l=6$ tetrahedra.

FCC tetrahedra form the basis for the construction of a number of the global structures observed in atomic clusters. For example, a perfect MacKay icosahedron is made up of 20 FCC tetrahedral units packed together so that they all share a single vertex at the center of the cluster, share three faces with other tetrahedra and have one face at the cluster surface. The decahedral structures are also constructed from tetrahedra. However, in both cases, the tetrahedra are not perfect and must be strained to form the complete spacing filling icosahedron or decahedron. The $l=5$ tetrahedra form the basis of the four layer MacKay icosahedron with 309 atoms and Fig.~\ref{fig:vmd} ($l=5$ MD) shows a cluster containing eight complete and several partially complete tetrahedra arranged to build an icosahedral core. The $l=6$ MD structure appears to be in the initial stages of forming the 591-atom MacKay icosahedron, which is the structure that forms the core of the lowest energy structure for our $N=600$ cluster.~\cite{xia204} However, the multiple tetrahedra in the $l=5$ and $l=6$ clusters are not always arranged to form partial icosahedra which suggests that they may lead to the formation of a variety of different core ordered clusters. The structures from the B minimum on the free energy surface contain marginally fewer completed $l=6$ tetrahedra and Fig.~\ref{fig:vmd} ($l=6$ MC) shows three connected subunits that could either be the initial formation of the top part of an icosahedron, or a partially complete decahedral core structure.

$l=7$ tetrahedra are needed to form a six-layer, $N=923$, icosahedron which is too large for the present system. Instead, we see the formation of clusters with a single tetrahedron dominating the core of the cluster with the vertices located just below the surface, similar to the Leary tetrahedron found at the core of the 98-atom LJ cluster global potential energy minimum.~\cite{lea90} We also see decahedral type structures with an $l=7$ chain of bulk icosahedral atoms running through the core. Both types of structure are observed in our free energy calculations and in our MD runs. The $l=8$ tetrahedron appears to be too large to fit in the core of the cluster and we only see partially completed structures with a single line of eight icosahedral atoms, leading to the decahedral type structures. These $l=8$ structures are located at the turning point between the negative and positive $Q_b - Q_a$ correlation in Fig.~\ref{fig:q6md} and, with increasing $Q_b$, we only see the formation of structures containing partially formed three dimensional arrangements of bulk icosahedral atoms. We do not see any chains of bulk icosahedral atoms longer than $l=8$. Eventually, the number of bulk icosahedral atoms decreases to zero, giving rise to pure FCC and HCP based structures.  

Noya and Doye~\cite{noya06} see an equivalent set of structures at intermediate temperatures in the 309-atom LJ cluster. The single tetrahedral core, multiply twinned tetrahedral core and ideal core icosahedron clusters are distinguished on the basis of their mean vibrational frequency which also provides a measure of the strain within the system. Clusters with larger but fewer tetrahedra are less strained and have higher vibrational frequencies. They also find that the multiple twinned tetrahedral structures are more commonly observed than the single tetrahedral core or perfect icosahedron. For the $N=600$ cluster studied here, we find that there is a small temperature range over which the free energy minimum associated with the single tetrahedral core phase is lower than that associated with the multiple twinned tetrahedral structures, which only appears at a lower $T$.  The appearance of a free energy minimum associated with this tetrahedron can be rationalized on the basis that the core, ordered into the {\it FCC} lattice, lowers the energy of the cluster, while the disordered surface provides a degree of entropic stability. As a result, we see a phase that is stable at intermediate temperatures but it is not possible to determine if the tetrahedral phase is the most stable state at any point over the temperature range studied because our calculations are carried out over a limited range in the bulk-surface order parameters. However, our MD simulations show that the phase is kinetically accessible, although at lower temperatures than we studied in our free energy calculations. We also note that the $l=5$ tetrahedra based structures are not seen in the free energy calculations and only appear at the lower temperatures studied with the MD.

\begin{figure}
\hbox to \hsize{\epsfxsize=1\hsize\hfil\epsfbox{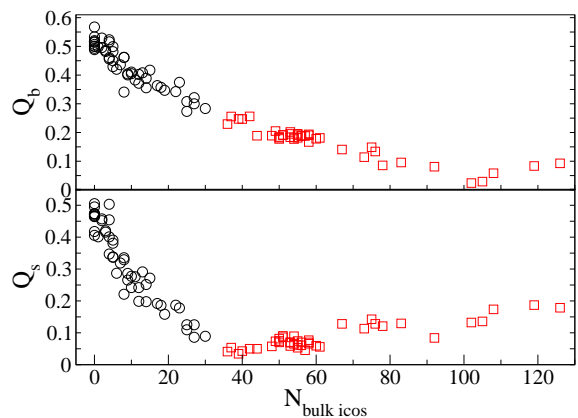}}
\caption{$Q_b$ (top) and $Q_s$ (bottom) as a function of the number of bulk icosahedral atoms, $N_{\rm bulk\: icos}$ for the quenched inherent structures obtained from the MD runs. The red and black circles represent structures from the negatively and positively correlated $Q_s - Q_b$ branches of Fig.~\ref{fig:q6md}, respectively. (In color online.) }
\label{fig:qbulk}
\end{figure}

The organization of the bulk icosahedral atoms into tetrahedral subunits may also help explain the structural correlations observed in Fig.~\ref{fig:q6md}. The presence of five-fold symmetric particles frustrates the bulk six-fold symmetry of the FCC lattice, leading to a decrease in $Q_b$ with increasing numbers of bulk icosahedral atoms ($N_{\rm bulk\: icos}$),  but  $N_{\rm bulk\: icos}$ also appears to be correlated with surface ordering even though these bulk atoms are not taken into account when calculating $Q_s$ (see Fig.~\ref{fig:qbulk}). In the absence of any icosahedral atoms, the surface order of a cluster is dominated by FCC facet formation which is then disrupted with the appearance a small number of isolated bulk icosahedral atoms. As $N_{\rm bulk\: icos}$ increases, these five-fold symmetric atoms begin to organize into linear chains that lead to the formation of decahedral type structures when the chain passes through the center or is just off the center. If the chain of bulk icosahedral atoms appears close to the surface, we see FCC based clusters with a defect plane of HCP atoms. These defects cause a further decrease in $Q_s$. However, for $N_{\rm bulk\: icos}>30$, the bulk icosahedral atoms are arranged in the three-dimensional tetrahedral units and we begin to see an increase in surface order. While it is difficult to definitively show, one possible explanation for the correlation is that the presence of the faces of the tetrahedra just below the surface act as an ordering template for the particles in the surface itself. As the number of tetrahedra increases, so does the ordering effect.

Understanding how clusters freeze to the different noncrystalline structures such as icosahedra or decahedra remains an important, unsolved problem. In the case of a bulk system freezing to a crystal, the critical nuclei that form are structurally related to the final solid structure and can form anywhere in the system because of the translational invariance of a crystal but an icosahedron or decahedron is a global structure and the local environments within the cluster appear very different. One of the challenges is to identify a suitable reaction coordinate for nucleation, which is an inherently local process, that can result in a globally organized structure. The free energy surfaces calculated in the present work provide us with information concerning the type of structures the cluster forms and the fact that the MD simulations freeze to the same structures that we observe in the MC studies supports the idea that these structures are important in the phase behavior of the clusters. However, $Q_b$ and $Q_s$ may not be good order parameters to describe the nucleation process because they suggest that, at low temperatures, the system needs to move through the configuration space of the central minimum, which is associated with structures containing multiply twinned $l=6$ tetrahedra, to form the single, core ordered, $l=7$ tetrahedral structure and this seems unlikely. Nevertheless, our work suggests that medium sized clusters have a propensity for building the tetrahedral subunits that form the basis of many of the noncrystalline structures. Understanding how the different sized tetrahedra are formed and how they are arranged within the core of a cluster may provide us with considerable insight into the nucleation problem in these systems.


\section{Conclusions}
Small Lennard-Jones clusters exhibit a variety of structural transformations including solid-solid transitions between the non-crystalline structures and surface reconstructions between MacKay and anti-MacKay icosahedra. Our work highlights the fact that, as the cluster size increases, structural transitions in the core become more important, especially at intermediate temperatures where the stability of a given phase results from a fine balance between energy and entropic contributions. In the present case, we find free energy minima associated with the formation of {\it FCC} tetrahedral subunits of different sizes in the core of the cluster with a disordered surface. This study also suggests that the nucleation, growth and rearrangement of the tetrahedral subunits may have an important role in the kinetic formation of the noncrystalline cluster phases such as icosahedra and decahedra.


\acknowledgments
We acknowledge NSERC and CFI for funding. Computing resources were provided by WESTGRID, ACEnet and SCARCNET.


\begin{thebibliography}{999}


\bibitem{bal05} F. Baletto and R. Ferrando, Rev. Mod. Phys. {\bf 77}, 371 (2005).

\bibitem{walesbook} D. J. Wales, {\it Energy Landscapes, With Applications to Clusters, Biomolecules and Glasses} (Cambridge University Press, Cambridge, 2003).

\bibitem{xia104} Y. H. Xiang, H. Y. Jiang, W. S. Cai and X. G. Shao, J. Phys. Chem. A {\bf 108}, 3586 (2004), and references therein.

\bibitem{xia204} Y. H. Xiang, L. J. Cheng, W. S. Cai and X. G. Shao, J. Phys. Chem. A {\bf 108}, 9516 (2004). 

\bibitem{temp} R. H. Swendsen and J.-S Wang, Phys. Rev. Lett. {\bf 57}, 2607 (1986).

\bibitem{lab90} P. Labastie and R. L. Whetten, Phys. Rev. Lett. {\bf 65}, 1567 (1990).
\bibitem{che92} H.-P. Cheng and R. S. Berry, Phys. Rev. A {\bf 45}, 7969 (1992).

\bibitem{doy99} J. P. K. Doye, M. A. Millar and D. J. Wales, J. Chem. Phys. {\bf 110}, 6896 (1999).
\bibitem{nei00} J. P. Neirotti, F. Calvo, D. L. Freeman and J. D. Doll, J. Chem. Phys. {\bf 112}, 10340 (2000).
\bibitem{cal00} F. Calvo, J. P. Neirotti, D. L. Freeman and J. D. Doll, J. Chem. Phys. {\bf 112}, 10350 (2000).


\bibitem{fra01} D. D. Frantz, J. Chem. Phys. {\bf 115}, 6136 (2001); J. Chem. Phys. {\bf 102}, 3747 (1995). 

\bibitem{mand06} V. A. Mandelshtam and P. A. Frantsuzov, J. Chem. Phys. {\bf 124}, 204511 (2006).

\bibitem{zhan07} L. Zhan, J. Z. Y. Chen and W.-K. Liu, J. Chem. Phys. {\bf 124}, 141101 (2007).
\bibitem{steinhardt} P. J. Steinhardt, D. R. Nelson, and M. Ronchetti, Phys. Rev. B {\bf 28}, 784 (1983).

\bibitem{doy02} J. P. K. Doye and F. Calvo, J. Chem. Phys. {\bf 116}, 8307 (2002).

\bibitem{noya06} E. G. Noya and J. P. K. Doye, J. Chem. Phys. {\bf 124}, 104503 (2006).

\bibitem{pol08} W. Polak, Phys. Rev. E {\bf 77}, 031404 (2008).

\bibitem{pol03} W. Polak and A. Patrykiejew, Phys. Rev. B {\bf 67}, 115401 (2003).


\bibitem{frenkel1996} P. R. ten Wolde, M. J. Ruiz-Montero, and D. Frenkel, J. Chem. Phys. {\bf 104}, 9932 (1996).

\bibitem{cone} Y. Wang, S. Teitel, and C. Dellago, J. Chem. Phys. {\bf 122}, 214722 (2005).

\bibitem{frenkelbook}
D. Frenkel and B. Smit, {\it Understanding Molecular Simulation: From Algorithms to Applications} 
(Academic Press, San Diego, 2002).

\bibitem{ih}
F. H. Stillinger and T. A. Weber, Phys. Rev. A 25, 978 (1982);
 {\it Science} {\bf 225}, 983-989 (1984);
F. H. Stillinger, {\it ibid} {\bf 267}, 1935-1939
(1995).


\bibitem{cla93} A. S. Clarke and H. Jonsson, Phys. Rev. E {\bf 47}, 3975 (1993).

\bibitem{hen01} S. C. Hendy and B. D. Hall, Phys. Rev. B {\bf 64}, 085425 (2001).

\bibitem{lea90} R. H. Leary and J. P. K. Doye, Phys. Rev. E {\bf 60}, R6320 (1990).






\end{thebibliography}
\end{document}